\documentclass[prd,twocolumn,showpacs]{revtex4-1}

\usepackage{amsfonts}
\usepackage{amsmath}
\usepackage{amssymb}
\usepackage{bm}
\usepackage{dcolumn}
\usepackage{graphicx}
\usepackage{graphics}
\usepackage[latin1]{inputenc}
\usepackage{latexsym}
\usepackage{rotating}
\usepackage{xspace} 
\usepackage[usenames]{color}
\usepackage{mathrsfs}

\usepackage{ulem}
\definecolor {darkgreen}{rgb}{0.2,0.7,0.2}

\newcommand\be{\begin{equation}}
\newcommand\ba{\begin{eqnarray}}
\newcommand\ee{\end{equation}}
\newcommand\ea{\end{eqnarray}}

\newcommand\bw{\begin{widetext}}
\newcommand\ew{\end{widetext}}

\newcommand{\ISCO}{{\mbox{\tiny ISCO}}}
\newcommand{\cont}{{\mbox{\tiny cont}}}

\newcommand{\fivePN}{{\mbox{\tiny 5PN}}}

\newcommand{\PP}{{\mbox{\tiny PP}}}
\newcommand{\FS}{{\mbox{\tiny FS}}}

\newcommand{\stat}{{\mbox{\tiny stat}}}
\newcommand{\sys}{{\mbox{\tiny sys}}}
\newcommand{\useful}{{\mbox{\tiny useful}}}
\newcommand{\fmax}{{\mbox{\tiny max}}}
\newcommand{\fmin}{{\mbox{\tiny min}}}
\newcommand{\inst}{{\mbox{\tiny inst}}}
\newcommand{\temp}{{\mbox{\tiny temp}}}
\newcommand{\true}{{\mbox{\tiny true}}}

\newcommand{\nPN}{{\mbox{\tiny nPN}}}
\newcommand{\PNfour}{{\mbox{\tiny 4PN}}}

\newcommand{\lambdabarnew}{\bar{\lambda}}

\newcommand{\mrm}{\mathrm}

\begin{document}
\title{Love can be Tough to Measure}

\author{Kent Yagi}
\affiliation{Department of Physics, Montana State University, Bozeman, MT 59717, USA.}

\author{Nicol\'as Yunes}
\affiliation{Department of Physics, Montana State University, Bozeman, MT 59717, USA.}

\date{\today}

\begin{abstract} 

The waveform phase for a neutron star binary can be split into point-particle terms and finite-size terms (characterized by the Love number) that account for equation of state effects. The latter first enter at 5 post-Newtonian (PN) order (i.e.~proportional to the tenth power of the orbital velocity), but the former are only known completely to 3.5 PN order, with higher order terms only known to leading-order in the mass-ratio. We here find that not including point-particle terms at 4PN order to leading- and first-order in the mass ratio in the template model can severely deteriorate our ability to measure the equation of state. This problem can be solved if one uses numerical waveforms once their own systematic errors are under control.

\end{abstract}



\maketitle

\section{Introduction}
One of the largest uncertainties in nuclear physics is the equation of state (EoS) at supra-nuclear densities. A neutron star (NS) is a perfect laboratory to study such physics. The NS mass-radius relation depends strongly on the EoS; an independent measurement of mass and radius (e.g.~with X-ray bursters and low-mass X-ray binaries) has led to a EoS constraint~\cite{lattimer-prakash-review,steiner-lattimer-brown,ozel-review,lattimer-review}. Future X-ray observations with e.g. NICER or LOFT may allow us to measure the NS mass and radius more precisely~\cite{Psaltis:2013fha}. Gravitational wave (GW) observations of NS binary inspiral with Adv.~LIGO, VIRGO and KAGRA, or future detectors such as LIGO-III and ET, may allow further constraints. As NSs inspiral, they deform each other through tidal interactions, which affect the orbital evolution, encoded in the waveform~\cite{flanagan-hinderer-love}.
 
The measurability of the EoS with GW observations depends on the measurement error. For such systems, parameter estimation is carried out through template-based likelihood analysis: the signal is cross-correlated with a template waveform, weighted by the spectral noise. The measurement error is then a combination of statistical error (due to detector noise) and systematic error (for example, due to waveform mismodeling). All previous EoS-related GW work only accounted for the former~\cite{flanagan-hinderer-love,read-love,hinderer-lackey-lang-read,lackey,damour-nagar-villain,lackey-kyutoku-spin-BHNS,read-matter,delpozzo,maselli-gualtieri-ferrari}, but the latter could dominate the error budget~\footnote{\cite{lackey-kyutoku-spin-BHNS} and~\cite{read-matter} estimate the systematic error due to errors in numerical relativity waveforms.}.

The binary NS waveform is the product of a slowly-evolving amplitude and a rapidly-varying phase; detectors are most sensitive to the latter. The phase is composed of point-particle terms (assuming the NSs have no internal structure) and finite-size terms (internal-structure corrections). Both of these are computed by expanding the Einstein equations in the ratio of the orbital velocity to the speed of light (a post-Newtonian (PN) expansion), where a term proportional to $(v/c)^{2N}$ is of $N$th PN order. Finite-size terms, characterized by the Love number, first enter at 5PN order~\cite{flanagan-hinderer-love}, but point-particle terms are only known completely up to 3.5PN order, with higher than 3.5PN order terms only known to leading-order in the mass-ratio. 

Are these PN expansions accurate filters to extract the EoS once a GW is detected from a NS binary inspiral? One may argue that the unknown mass-ratio corrections to the point-particle terms at 4PN order and higher will be smaller than the finite-size terms; the latter are multiplied by an inverse power of the NS compactness, which leads to a large coefficient of ${\cal{O}}(10^{3})$. Although these unknown terms seem comparatively small, we will here explicitly show that not including them can destroy the accuracy to which the EoS can be measured.

\section{Gravitational Waveform Phase}

Consider a NS binary with component masses $m_A$ and radii $R_A$ [$A=(1,2)$]. Its Fourier GW phase $\Psi(f)$, can be written as a linear combination of a point-particle contribution $\Psi^{\PP}(f)$ and a finite-size contribution $\Psi^{\FS}(f)$. 

$\Psi^{\PP} (f)$ assumes the NSs are test-masses with no internal structure; it is completely known to 3.5PN order~\cite{arun35PN}, with higher order terms known only to leading-order in the symmetric mass ratio $\eta \equiv m_1 m_2/M^2$, where $M \equiv m_1 + m_2$~\cite{fujita-22PN,bardeen}. The leading-order-in-$\eta$ 4PN term is in~\cite{varma}, while the 5PN term is 
\be
\label{eq:psi5-PP}
\Psi_\fivePN^{\PP} = ( c_1 + c_2 \ln \sqrt{x} ) \frac{x^{5/2}}{\eta}\,, 
\ee
where $x \equiv (\pi M f)^{2/3}$, with $f$ the GW frequency, and $(c_1, c_2) \approx (-210,5.5)$ pure numbers.

The above point-particle contributions neglect the NS's internal structure, which will also affect the waveform phase. This structure is encoded in the $\ell$-th electric ($\lambda_{\ell\geq2}$) and magnetic ($\sigma_{\ell\geq2}$) tidal deformability parameters~\cite{damour-nagar,binnington-poisson}. These parameters are defined by the ratio of the induced $\ell$-th mass or current multipole moment to the $\ell$-th electric or magnetic tidal tensor. They represent the susceptibility of a NS of being deformed by an external tidal force. The electric deformability $\lambda_\ell$ is related to the $\ell$-th electric tidal Love number, $k_\ell \equiv [(2 \ell - 1)!!/2] \lambda_\ell/R^{2\ell +1}$, which is the second apsidal constant in the Newtonian limit~\cite{damour-nagar,kent-multipole-love}. 

The NS's internal structure enters the waveform first at 5PN order through the $\ell=2$ electric deformability~\cite{flanagan-hinderer-love}:
\be
\label{FS-effect}
\Psi^{\FS}_\fivePN(f) = -\frac{3}{8} \frac{1}{\eta} \sum_{A=1,B\neq A}^{2} \left( \frac{R_A}{M} \right)^5 \left( 1+12 \frac{m_B}{m_A} \right) k_{2}^{(A)} x^{5/2}\,.
\ee
PN corrections can be found in~\cite{damour-nagar-villain}, while $\lambda_{\ell\geq3}$ and $\sigma_{\ell\geq2}$ enter first at 7 and 6PN order respectively~\cite{kent-multipole-love}. Notice that the leading PN order, finite-size terms depend on $k_{2}^{A} R_{A}^{5} \propto \lambda_{2}^{A}$, which in turn depends on the NS EoS.

\section{Useful GW Cycles}

The amount of information in each phase term relative to the detector's noise will determine the accuracy to which that given term can be extracted. One estimate of this are the useful GW cycles~\cite{damour-useful}, roughly the number of cycles contained in a given phase term weighted by the noise:   
\be
\label{eq:useful}
N_\mrm{useful} = \left( \int^{f_{\fmax}}_{f_{\fmin}} \frac{df}{f} w(f) N_\inst(f) \right) \left( \int^{f_{\fmax}}_{f_{\fmin}} \frac{df}{f} w(f)  \right)^{-1}\,,
\ee
where $w(f) = A[t(f)]^2/[f S_n(f)]$, with $S_n(f)$ the detector's spectral noise and $A(t)$ the time-domain waveform amplitude, $(f_\fmin,f_\fmax)$ are the minimum and maximum GW frequencies during an observation period, and $N_\inst(f) \equiv f^2/\dot{f}$, with $\dot{f} = 2 \pi (d ^2 \Psi/d f^2)^{-1}$~\cite{kent-multipole-love} and $\Psi(f)$ the given phase term. The \emph{instantaneous} number of GW cycles $N_\inst(f)$ is related to the total number of cycles by $N = \int^{f_{\fmax}}_{f_{\fmin}} (N_\inst/f) df$. The latter is not weighted by the spectral noise, and thus, it is {\emph{not}} a robust measure of the amount of information in a given phase term.

\begin{figure}[htb]
\begin{center}
\includegraphics[width=8.5cm,clip=true]{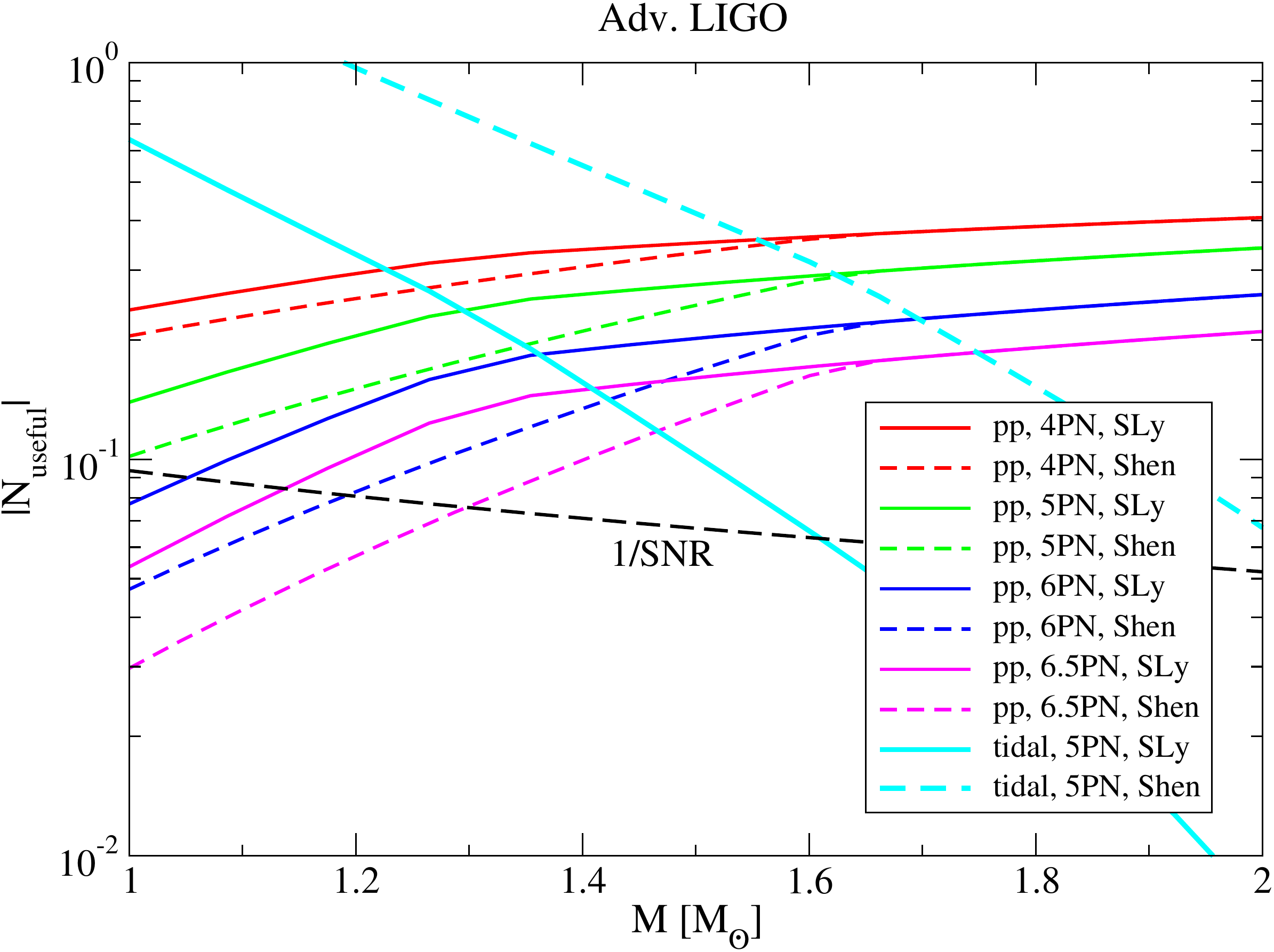}  
\caption{\label{fig:useful}(Color online) $N_\useful$ vs. mass for point-particle phase terms to leading order in $\eta$ and for the leading-order, finite-size GW phase term. We focus on an Adv.~LIGO detection, with (high-power, zero-detuned) spectral noise~\cite{AdvLIGO-noise}, $f_\fmin = 10$Hz and $f_\fmax = \min(f_\ISCO,f_\cont)$, with $f_{\ISCO} = (6^{3/2} \pi M)^{-1}$ the innermost-stable circular orbit frequency for a point-particle in a Schwarzschild background and $f_{\cont}$ the approximate contact frequency. The finite-size terms are modeled with two representative EoSs for realistic NSs (SLy~\cite{SLy}, Shen~\cite{Shen1,Shen2}), both of which allow for stars above the PSR J0348+0432 limit~\cite{2.01NS}. Other realistic EoSs~\cite{APR,LS} lead to results that fall between those shown. We also show 1/SNR for a NS binary at luminosity distance $D_L=100$Mpc. Observe that the finite-size terms and the (incomplete) point-particle terms lead to a comparable number of useful cycles, all above the rough 1/SNR threshold.
}
\end{center}
\end{figure}
Figure~\ref{fig:useful} shows $N_\useful$ as a function of mass, for the leading-PN-order, finite-size phase term and leading-order-in-$\eta$, point-particle terms. Roughly speaking, a given phase term affects parameter estimation if its useful number of cycles are above the inverse of the signal-to-noise ratio (SNR)~\cite{cornish-PPE}, since the phase measurement accuracy is roughly 1/SNR. Observe that the useful number of cycles is comparable for point-particle and finite-size terms, both of which are generally above 1/SNR. For low masses, $N_\useful$ depends on the EoS because the high-frequency cutoff is the contact frequency. This shows that the terms in the phase in question have a large contribution in the high-frequency regime. For high masses, the useful cycles become independent of the EoS because the contact frequency exceeds $f_\ISCO,$ which does not depend on the NS internal structure.

These results contradict the prior belief that point-particle terms at high PN order are negligible when compared to leading-order finite-size terms~\cite{flanagan-hinderer-love}. That belief is rooted in that the latter are enhanced relative to the former by a factor of $C_{A}^{-5} \sim 10^5$, where $C_{A} \equiv m_{A}/R_{A}$ is the NS compactness. Equation~(\ref{FS-effect}), however, shows that the leading-order, finite-size term is actually proportional to $(R/M)^5 \sim 10^3 \ll (R_{A}/m_{A})^{5}$, while Eq.~(\ref{eq:psi5-PP}) shows that the 5PN point-particle term has a large coefficient  $c_1/\eta \sim \mathcal{O}(10^3)$. One can then see analytically that $\Psi^\FS_\fivePN$ and $\Psi^\PP_\fivePN$ will lead to comparable contributions to the phase: $\Psi^\PP_\fivePN/\Psi^\FS_\fivePN \sim 6 \left( k_2/0.1 \right)  \left( C_{A}/0.1 \right)^{-5}$; the point-particle contribution is generically larger than the finite-size one at 5PN order. 

\section{Systematic vs. Statistical Errors}

Just because the point-particle and finite-size phase terms contribute similarly to the total phase need not imply that not including them deteriorates the EoS measurement. To study this, let us now compare estimates of the statistical error (due to random detector noise) and systematic error (due to not including point-particle terms at 4PN order and higher) on the extraction of $\lambda_{2}$.

The statistical error on the extraction of parameter $\theta^{i}$ can be roughly estimated to be
\be
\label{eq:statisical}
\Delta_\stat \theta^i = \sqrt{(\Gamma^{-1})_{ii}}\,,
\ee
(no Einstein summation implied). $\Gamma_{ij} \equiv (\partial h_\temp/\partial \theta^i | \partial h_\temp/\partial \theta^j)$ is the Fisher information matrix, $h_\temp$ the waveform template and $(A|B)$ the noise-weighted inner product~\cite{cutlerflanagan}. This estimate applies only to signals in Gaussian, stationary noise at high SNR and assuming the template matches the signal perfectly~\cite{vallisneri-fisher, cutler-vallisneri}. One can interpret $\Delta_\stat \theta^i$ as the width of the posterior distribution of the recovered $\theta^{i}$, which would be obtained through a Bayesian inference study.  

Since PN templates are approximate solutions to the Einstein equations, they will always be contaminated by systematic mismodeling error. This error is roughly~\cite{cutler-vallisneri}
\be
\label{eq:systematic}
\Delta_\sys \theta^i = \left( \Gamma^{-1} \right)^{ij} \left( [ h_{\true} - h_{\temp} ] | \partial_j h_{\temp} \right)\,, 
\ee
where $h_{\true}$ is the signal and $h_{\temp}$ is the template. This estimate assumes signals in Gaussian, stationary noise at high SNR. One can associate it with a shift in the peak of the posterior distribution of the recovered $\theta^{i}$, which again could be better estimated with a Bayesian analysis. Within the approximations considered, statistical errors are proportional to the inverse of the SNR, while systematic errors do not depend on the SNR, although they both depend on the shape of the noise curve. There always exists a sufficiently high SNR where systematic errors dominate the error budget.   

But are the SNRs we expect with second- and third-generation GW detectors so high? Consider equal-mass, non-spinning NS binary, quasi-circular inspirals. The true signal and the templates will be parameterized by $\theta^i = (\ln \mathcal{M}, \ln \eta, t_c, \phi_c, \ln D_L, \lambdabarnew_{2,s})$, where $\mathcal{M} \equiv M \eta^{3/5}$ is the chirp mass, $t_c$ and $\phi_c$ are the time and phase at coalescence, while $\lambdabarnew_{2,s} \equiv ( \lambdabarnew_{2}^{(1)} + \lambdabarnew_{2}^{(2)})/2$ is the averaged, dimensionless tidal deformability, with $\lambdabarnew_2^{(A)} \equiv \lambda_2^{(A)}/m_A^5$. For the SLy and Shen EoSs, $\lambdabarnew_2^{(A)}\approx277$ and $1212$ respectively~\cite{I-Love-Q-PRD}. We concentrate on equal-mass signals because this minimizes the statistical error on $\lambdabarnew_{2}^{(A)}$~\footnote{For an unequal-mass system with mass difference $|m_1 - m_2| \gtrsim 0.2 M_\odot$, one also needs to include $\lambdabarnew_{2,a} \equiv ( \lambdabarnew_{2}^{(1)} - \lambdabarnew_{2}^{(2)} )/2$ in the parameter list~\cite{kent-multipole-love}. This increases the dimensionality of the Fisher matrix, which induces a larger statistical error on $\lambdabarnew_{2,s}$, due to strong correlation with $\lambdabarnew_{2,a}$.}, thus providing the best hope to measure the NS EoS. We evaluate Eq.~(\ref{eq:systematic}) at the best-fit values $t_c = 0 = \phi_c$ and $D_L=100$Mpc, for SNRs in $(10,20)$ for Adv.~LIGO.

The Fourier transform of the template $\tilde{h}_\temp (f)$ will be modeled in the restricted PN approximation with the stationary-phase approximation~\cite{cutlerflanagan}, including up to 3.5PN order~\cite{arun35PN} terms in $\Psi^\PP$ and up to 7.5PN order corrections to the $\Psi^\FS$ in Eq.~(\ref{FS-effect}) ($\lambda_{2}$ terms only)~\cite{damour-nagar-villain}. The Fourier transform of the signal will be modeled via
\be
\tilde{h}_\true (f) = \tilde{h}_\temp (f) \exp[i \Psi^\PP_\nPN(f)]\,,
\ee
where $\Psi^\PP_\nPN(f)$ is the $n$th PN order, point-particle phase term. The true signal and the template differ only due to $\Psi^\PP_\nPN(f)$. We will estimate the systematic errors on $\theta^i$ due to not including $\Psi^\PP_\nPN(f)$ in the waveform template.  

\begin{figure}[t]
\begin{center}
\includegraphics[width=8.5cm,clip=true]{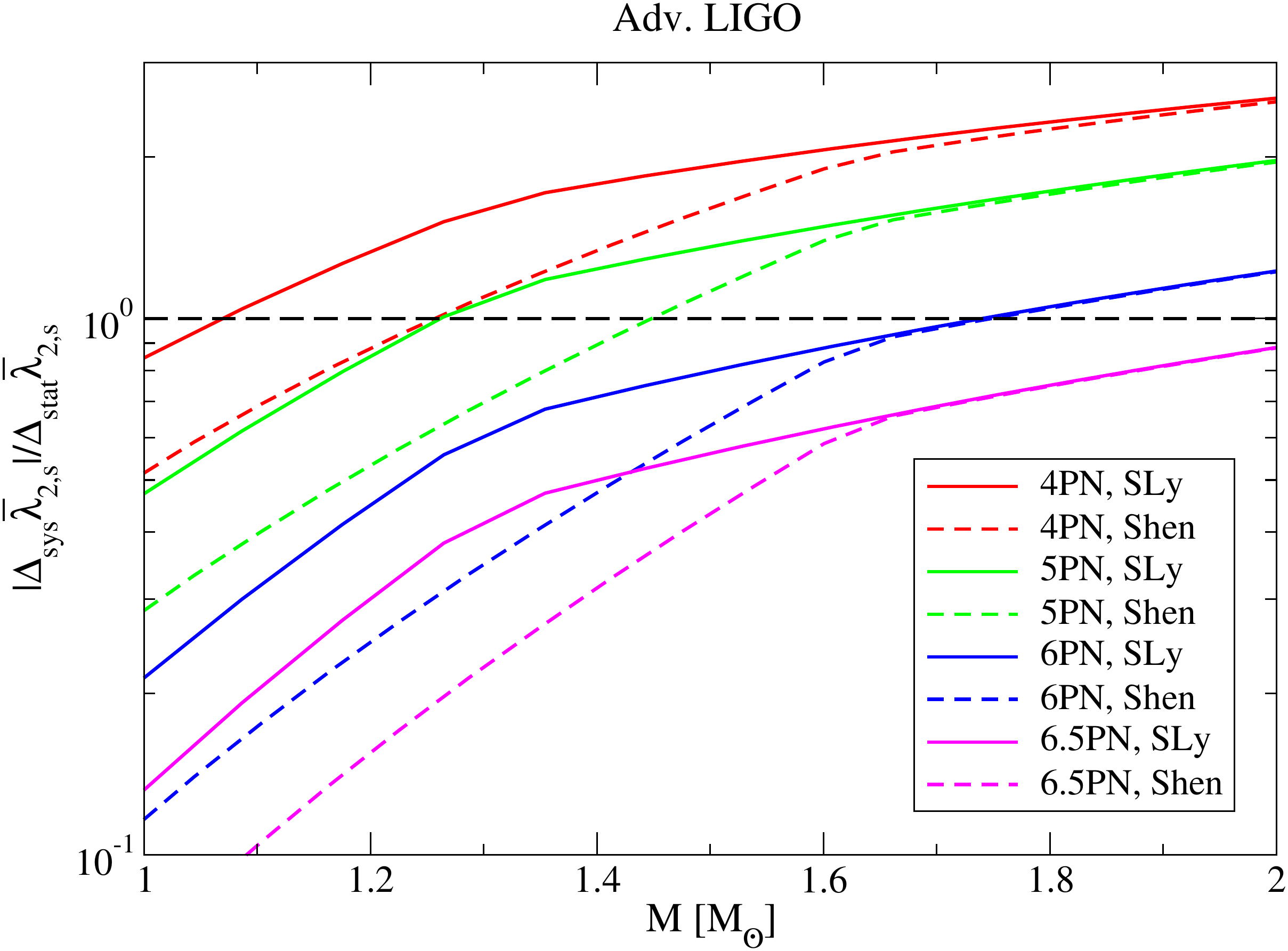}  
\caption{\label{fig:sys-eta0}(Color online) Ratio of the estimate of the systematic to the statistical error on the averaged dimensionless deformability $\lambdabarnew_{2,s}$ versus NS mass for Adv.~LIGO. The systematic errors arise due to not including the $n$th PN order point-particle term to leading-order in $\eta$, which is currently known. The statistical error is induced by detector noise. These errors are estimated using the NS EoSs SLy and Shen, as explained in Fig.~\ref{fig:useful}. 
Observe that the systematic error dominates the error budget when $n \leq 6$.
}
\end{center}
\end{figure}
Figure~\ref{fig:sys-eta0} shows the ratio of the systematic to the statistical errors as a function of NS mass for Adv.~LIGO, using signals with different $n$ and to leading-order in the mass ratio. The systematic errors dominate the statistical ones, unless one includes up to 6PN order, point-particle terms in the template. The importance of the systematic errors grows with increasing NS mass because the difference between the point-particle and the finite-size terms also grows with mass, as shown in Fig.~\ref{fig:useful}. For high masses, the ratio becomes independent of the EoS, as in Fig.~\ref{fig:useful}, since the integrals are truncated at $f_{\ISCO}$. The ratio does not depend on the EoS even in the high mass regime because $\partial \tilde{h}/\partial \lambdabarnew_{2,s}$ in Eqs.~(\ref{eq:statisical}) and~(\ref{eq:systematic}) has a $\lambdabarnew_{2,s}$-dependence only in the phase, and this cancels when computing the correlation.

We can confirm these results with an order of magnitude estimate. From Eq.~(\ref{eq:systematic}), the systematic error due to neglecting the 5PN, point-particle term in the phase is 
\be
\left( \Delta_\sys \ln \lambdabarnew_{2,s} \right)_\fivePN \approx (\Gamma^{-1})^{\lambdabarnew_{2,s} j} \frac{\Psi^\PP_\fivePN}{\Psi^\FS_\fivePN}  \Gamma_{j \lambdabarnew_{2,s}}  \approx 0.9 \left( \frac{500}{\lambdabarnew_{2,s}} \right) 
\ee
where in the last equality the Fisher matrices canceled and we evaluated the result for an equal-mass NS binary. We will show later that $\Delta_\stat \ln \lambdabarnew_{2,s} = \mathcal{O}(1)$, and thus, $(\Delta_\sys \lambdabarnew_{2,s})_\fivePN/\Delta_\stat \lambdabarnew_{2,s} = \mathcal{O}(1)$, as shown in Fig.~\ref{fig:sys-eta0}.

One can avoid introducing the above systematic error by including the leading-order-in-$\eta$, point-particle terms in the GW phase, but is the result accurate enough to extract the EoS?  This question cannot be formally answered because of our ignorance of the higher-order-in-$\eta$ terms in the point-particle phase. Nonetheless, one can determine what the magnitude of the coefficients of such terms has to be in order for the systematic error induced by not including them to be smaller than the statistical error. Decompose the $n$th PN order, point-particle term $\Psi^\PP_\nPN(f)$ into 
\be
\Psi^\PP_\nPN(f) = \sum_{k=0} \psi^\PP_{\nPN,k} \; \eta^{k-1} \; x^{(-5+n)/2}\,,
\ee
and estimate the systematic error $\Delta_\sys \lambdabarnew_{2,s}$ induced by not including the $k$th term in $\Psi^\PP_\nPN$ in the template model. 

\begin{figure}[t]
\begin{center}
\includegraphics[width=8.5cm,clip=true]{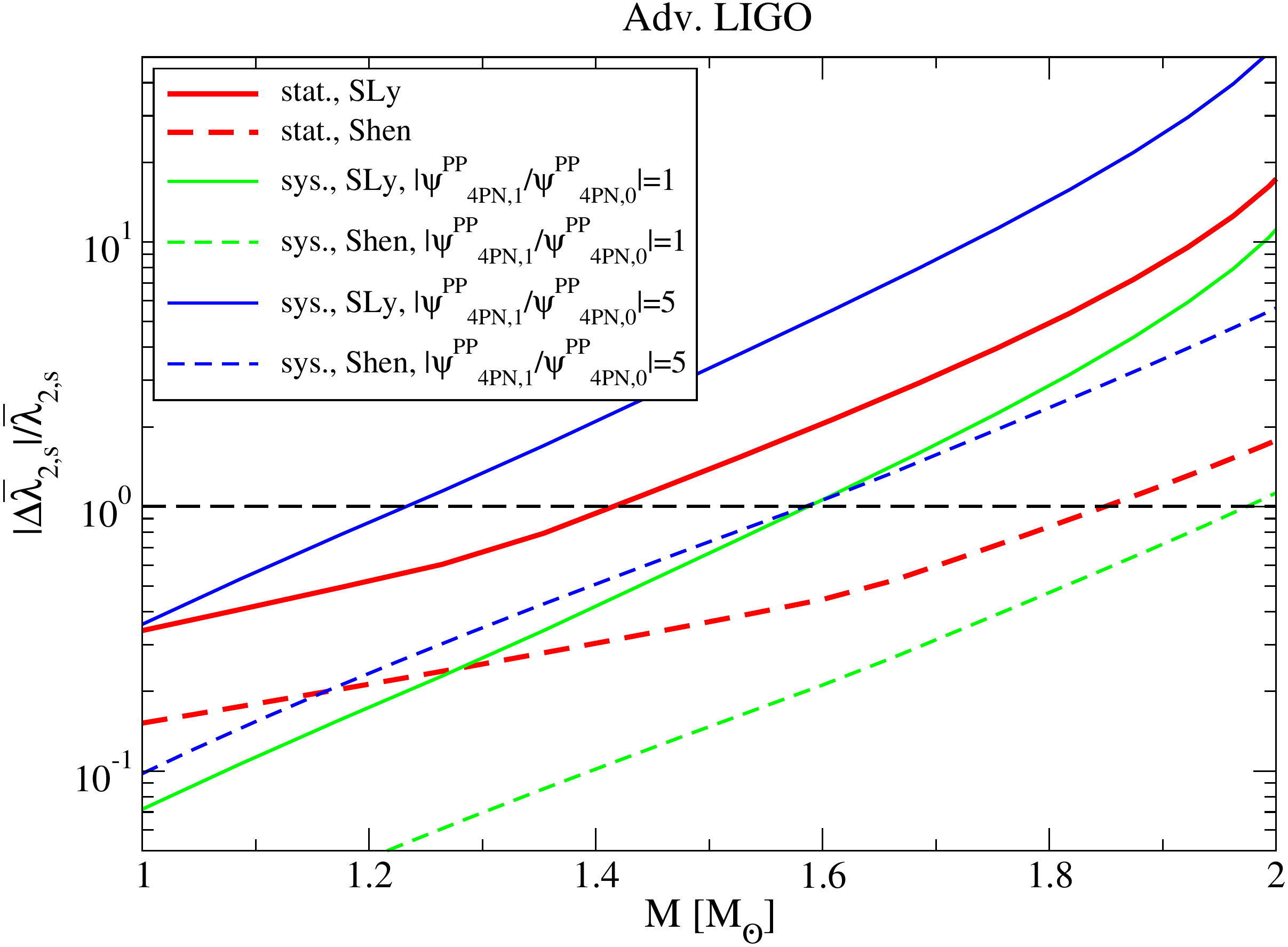}  
\caption{\label{fig:sys-stat}(Color online) Statistical and systematic errors on $\ln \bar{\lambda}_{2,s}$ due to not including the 4PN term at next-to-leading order in $\eta$ (currently unknown) for SLy and Shen EoSs using Adv.~LIGO. For the latter, we set $|\psi^\PP_{\PNfour,1}/\psi^\PP_{\PNfour,0}|=1$ and 5, while other choices can be obtained through a simple linear rescaling. Observe that when the coefficient of the next-to-leading order term is large enough, the systematic error dominates over the statistical one.
}
\end{center}
\end{figure}
Figure~\ref{fig:sys-stat} shows the statistical and systematic errors in the $n=4$ case with the SLy and Shen EoSs using Adv.~LIGO. The statistical error is consistent with the findings of~\cite{hinderer-lackey-lang-read,flanagan-hinderer-love,read-matter,damour-nagar-villain}. The systematic error in Fig.~\ref{fig:sys-stat} is calculated by choosing $|\psi^\PP_{\PNfour,1}/\psi^\PP_{\PNfour,0}| = 1$ and $5$. If the statistical error dominates the error budget, one can measure $\bar{\lambda}_{2,s}$ only if $m_1 = m_{2} \lesssim 1.4M_\odot$ for a SLy EoS, but $m_1 = m_{2} \lesssim 1.8M_\odot$ for a Shen EoS, which is again consistent with~\cite{hinderer-lackey-lang-read,flanagan-hinderer-love,read-matter,damour-nagar-villain}. We see that when $|\psi^\PP_{\PNfour,1}/\psi^\PP_{\PNfour,0}| = 5$, the statistical error is smaller than the systematic one for all masses when using a SLy EoS and for $m_{1} = m_{2} \gtrsim 1.15 M_{\odot}$ when using a Shen EoS. In such a case, the error budget is much larger than what was previously estimated in~\cite{hinderer-lackey-lang-read,flanagan-hinderer-love,read-matter,damour-nagar-villain} with a purely statistical analysis.

Are the choices made in Fig.~\ref{fig:sys-stat} for $|\psi^\PP_{\PNfour,1}/\psi^\PP_{\PNfour,0}|$ realistic? Let us estimate the ratio $|\psi^\PP_{\nPN,1}/\psi^\PP_{\nPN,0}|$ for $n \in (0,3.5)$, since the point-particle phase terms are completely known up to 3.5PN order. This ratio is in the range $(0.141,11.9)$, and thus, the choices made in Fig.~\ref{fig:sys-stat} are close to the mean. If $|\psi^\PP_{\PNfour,1}/\psi^\PP_{\PNfour,0}|$ is close to the maximum (minimum) of this range, then the systematic error would be dominant (subdominant) with respect to the statistical error, as shown in Fig.~\ref{fig:sys-stat}.

Perhaps a better estimate of this ratio can be obtained by approximating $\psi^\PP_{\PNfour,1}$ from the known lower PN order terms. One can take the 3PN binding energy, Kepler's law and the 3.5PN energy flux expression~\cite{blanchet-review}, invert them to calculate the GW phase, and keep the 4PN terms in the Taylor expansion to find a partial and incomplete expression for $\Psi^\PP_{\PNfour}$. The incompleteness is because this does not account for 4PN corrections to the binding energy, Kepler's law or the energy flux, since they are unknown. Doing so, $|\psi^\PP_{\PNfour,1}/\psi^\PP_{\PNfour,0}| \sim 10.3$, which is close to the maximum discussed above. For such a high ratio, the systematic error due to neglecting the next-to-leading order term in $\eta$ dominates the error budget. 

The results found here depend strongly on the detector considered. For an initial Adv.~LIGO configuration (no-SRM), the statistical error will be generally higher than the systematic one. But in this case, the fractional statistical error itself is above unity (except for very stiff EoSs at very low masses), and the EoS is not measurable. For third-generation detectors, the systematic error overwhelms the statistical one because the latter scales with the inverse of the SNR, while the former is independent of it. For the systematic error to be smaller than the statistical one for all NS masses, $|\psi^\PP_{\PNfour,1}/\psi^\PP_{\PNfour,0}| \lesssim 0.4$ for a hypothetical LIGO-III detector~\cite{LIGO3-noise} and $\lesssim 0.1$ for ET~\cite{ET-noise}. This is close to the minimum of the range of $|\psi^\PP_{\nPN,1}/\psi^\PP_{\nPN,0}|$ discussed above. Thus, third-generation detectors require more accurate modeling to control systematic mismodeling error.

\section{Discussion}

We have studied whether a wide class of waveform templates are sufficiently accurate to extract the EoS. We found three main results: (i) the point-particle phase terms at 4PN order and higher contribute to the noise-weighted cycles as much as finite size phase terms at 5PN order, thus contributing equally to parameter estimation; (ii) not including the leading-order-in-$\eta$ point-particle phase terms in the template model introduces a systematic error that dominates the error budget; (iii) the inclusion of these leading-order terms in the template is not sufficient to control the systematic error, as neglecting the next-to-leading-order-in-$\eta$ terms at 4PN order also introduces large systematic errors. 

Our results\footnote{After completing this paper, we were made aware that a broader analysis was independently and simultaneously being carried out~\cite{favata-sys}. We have compared our results with theirs~\cite{favata-sys}, and find that (after correcting an error in~\cite{favata-sys}) our results were in good agreement with theirs.  Our results thus quantitatively agree with those of~\cite{favata-sys} in the appropriate limits.  The major difference is that we concentrate on EoS measurements, and go beyond their analysis for that problem.}
 suggest that if one wishes to prevent systematic errors from contaminating EoS measurements, one may have to include the next-to-leading-order-terms-in-$\eta$ in the point-particle phase at least to 4PN order, either through direct PN calculation, through the construction of a hybrid template matched to numerical relativity results~\cite{boyle,lackey,lackey-kyutoku-spin-BHNS,read-matter}, or through resummation of lower-order PN terms that match numerical relativity results~\cite{EOB-review1,EOB-review2,damour-nagar-bernuzzi,hotokezaka2}. Numerical waveforms have their own systematic errors that also need to be under control. 

{\emph{Acknowledgments}}.~We would like to thank Katerina Chatziioannou, Neil Cornish, Eanna Flanagan, Tanja Hinderer, Scott Hughes and Leo Stein for useful comments, suggestions and advice. We also thank Marc Favata for reading our manuscript and making suggestions. N.Y. acknowledges support from NSF grant PHY-1114374, NSF CAREER Grant PHY-1250636 and NASA grant NNX11AI49G. Some calculations used the computer algebra-systems MAPLE, in combination with the GRTENSORII package~\cite{grtensor}.

\bibliography{master}
\end{document}